\documentclass[prl,twocolumn,showpacs,superscriptaddress,amsmath,amssymb]{revtex4-1}
\usepackage{graphicx}
\usepackage{bm}
\usepackage[dvips]{color}
\begin{document}
\title{On the Equilibrium Shape of Erythrocytes}
\author{V.I. Marchenko and E.R. Podolyak\\ P.L. Kapitza Institute for Physical Problems RAS, 119334 Moscow, Russia}

\begin{abstract}
It is shown that the model proposed by Canham quantitatively
describes the observed biconcave shape of human erythrocytes.

DOI: 10.1134/S1063776115040226
\end{abstract}



\maketitle
The first quantitative data on the biconcave shape of erythrocytes were obtained by Ponder in 1930 \cite{P30}. In 1972, Canham showed in [2] that such a shape can be obtained as
a result of minimization of energy, \begin{equation}\label{A}A{\int}H^2dS,\end{equation}
where $H$ is the average curvature of the membrane bounding an erythrocyte for a given volume $V$ and surface area $S.$ In 1976, Deuling and Helfrish \cite{DH76} proposed that energy (1) be
supplemented with a linear (in curvature) term,
\begin{equation}\label{B}B{\int}HdS,\end{equation}
occurring due to the difference between the media inside and outside an erythrocyte. Later, this theory was complicated when the authors tried to take into account nonlocal interaction between membranes and the shear elasticity of membranes and cytoskeleton in
an erythrocyte (see, for example, \cite{MLW}). As a result, the number of material parameters introduced for describing the shape of erythrocytes became too large.

In our opinion, the simplest model (1) was rejected due to inappropriate comparison of the theory with experimental data. In this comparison, so-called modified Cassini ovals were used, which are parametrized by three constants. For a fixed volume and surface area, only one parameter specifying the shape is left. This is admissible for demonstration purposes, but is insufficient for a substantiated conclusion about the applicability of the theory. In addition, in view of
the nonlinearity of the equilibrium equation, it is meaningless to compare the theory with a certain average shape for an erythrocyte group [5] when the spread of parameters in the group is significant (up to $20\%$).

In this communication, we focus on the fact that the Canham model [2] describes the observed shape quantitatively without any fitting parameters.

The variational problem discussed here has been comprehensively studied [6]. However, since an analytic solution is not available, numerical integration should be carried out for comparison with experimental data.

The result shows that it is impossible to single out the effect of spontaneous curvature in view of experimental errors (this obviously means that the characteristic sizes of an erythrocyte are very small as compared to ratio $A/B$). For this reason, we omit the relevant contribution (2). We set the modulus of $A,$ which is positive in accordance with the stability condition, equal to unity. Taking into account the constancy of the volume and surface area and using the method of Lagrangian multipliers, we have the
variational problem
\begin{equation}\label{E}
\delta\left({\int}H^2ds-\lambda_vV-\lambda_sS\right)=0.
\end{equation}

The Lagrangian multipliers $\lambda_v,\,\lambda_s$  have the following physical meaning [3]: $\lambda_v$ -- is the difference between external and internal pressures and $\lambda_s$ -- is the surface tension of the membrane. Both these parameters are small in view of the smallness of the effects of curvature (1) for macroscopic sizes of erythrocytes noticeably exceeding molecular size; the values of these parameters are chosen so that the observed volume and surface area are ensured.

The axisymmetric shape is set by the function ${z = f(r);}$ in this case, the average curvature is given by
\begin{equation}\label{H}
H=\frac{1}{r}\left(\frac{rf'}{\sqrt{1+f'^2}}\right)',
\end{equation}
where a prime indicates differentiation with respect to radius $r.$ Then problem (3) is reduced to the problem
\begin{equation}\label{H}
H=\frac{1}{r}\left(\frac{rf'}{\sqrt{1+f'^2}}\right)'
\end{equation}

The corresponding variational equation has the first integral
\begin{equation*}
2\frac{r\left(h\sqrt{1+f'^2}\right)'}{\left(1+f'^2\right)^{3/2}} -\frac{r(h^2-\lambda_s)f'}{\sqrt{1+f'^2}}-\frac{\lambda_v}{2}r^2=C.
\end{equation*}

Function $f$ has no singularity at ${r=0}$ only if integration constant $C$ vanishes.
Further integration was carried out numerically by the shooting method. We started from the neighborhood of the maximal radius (which was chosen as the unit of length), where ${f\propto\sqrt{1-r},}$  and tried to ensure the condition ${f'=0}$ for ${r\rightarrow0}.$

Let us introduce the parameters of length, specifying the volume ${V=(4\pi/3)R^3_v}$ and the surface area ${S=4{\pi}R^2_s}.$  The problem under investigation is meaningful if the reduced volume parameter ${v=(R_v/R_s)^3}$ is smaller than unity. In accordance with the results of general analysis, the biconcave shape corresponds to the absolute minimum of energy (1) in a narrow interval of the reduced volume: ${0.592<v<0.651}$ (see\,\cite{SBL91},\,Fig.\,9). Erythrocytes are precisely in this interval (see\,figure).

\begin{figure}\begin{center}
\includegraphics[width=\columnwidth]{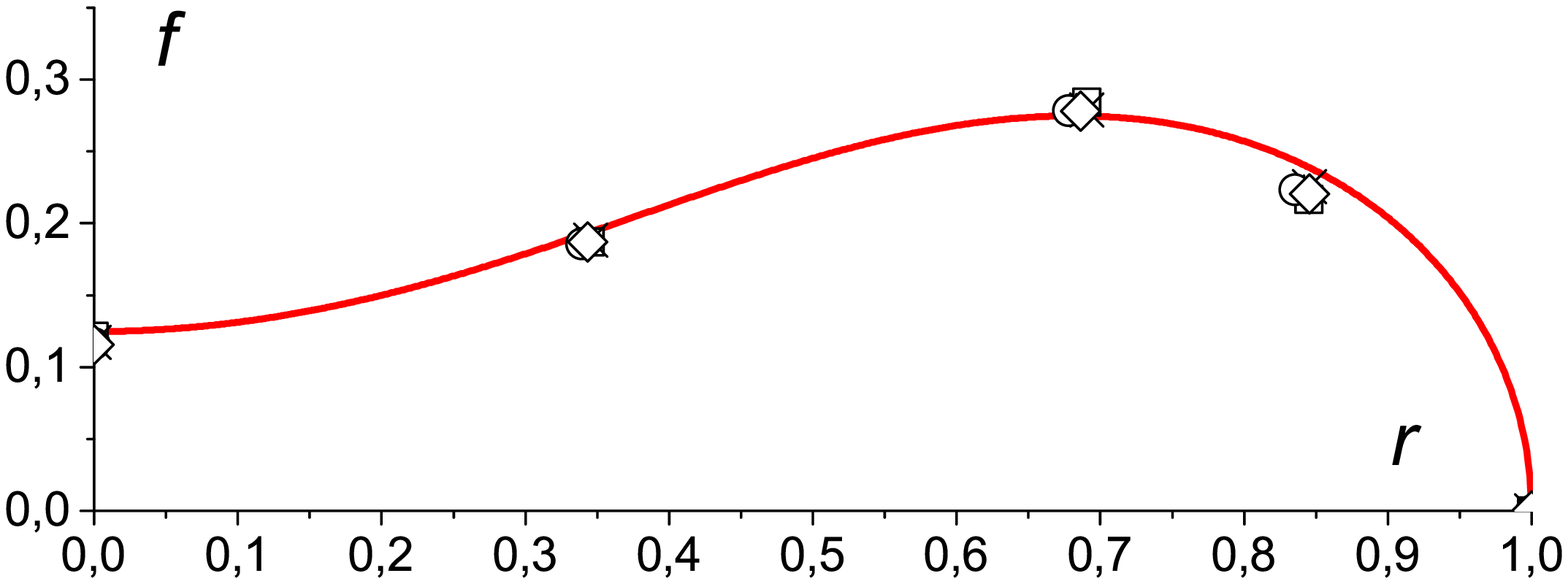}
\caption{Result of comparison of the theoretical shape for $v=0.610$ with the data obtained by Ponder [1] for four very close (in size) erythrocytes of human blood.}
\end{center}\end{figure}

Thus, the Canham model provides a quantitative description of the biconcave shape of an erythrocyte, and only very sound experimental arguments are required for the rejection of this remarkable theory. It should be noted that the cup-shaped form of erythrocytes, an alternative to the biconcave shape, is also obtained \cite{SBL91} when the problem under investigation is solved disregarding spontaneous curvature for ${v<0.592}.$ However, we are not aware of experimental data suitable for rigorous comparison with the theory, since shapes distorted by thermal fluctuations are usually observed.

\,

\,

JETP 120(4), 751 (2015) 

Original Russian Text - ZhETF 147(4), 867 (2015)

Received December 30, 2014

mar@kapitza.ras.ru

Translated by N. Wadhwa

\end{document}